\documentclass[pre,a4paper,superscriptaddress,showpacs,nofootinbib]{revtex4}

\usepackage{amsmath}    
\usepackage{graphicx}   
\usepackage{float}
\usepackage{subfigure}

\begin{document}

\title{Atomistic Simulations of Elastic and Plastic Properties in Amorphous Silicon}
\author{M. Talati*}
\affiliation{Universit\'{e} de Lyon, F-69622, France\\ Univ. Lyon 1, Laboratoire PMCN; CNRS, UMR 5586 F-69622 Villeurbanne Cedex}
\email[Email: ]{mtalati@lpmcn.univ-lyon1.fr}   
\author{T. Albaret}
\affiliation{Universit\'{e} de Lyon, F-69622, France\\ Univ. Lyon 1, Laboratoire PMCN; CNRS, UMR 5586 F-69622 Villeurbanne Cedex}
\author{A. Tanguy}
\affiliation{Universit\'{e} de Lyon, F-69622, France\\ Univ. Lyon 1, Laboratoire PMCN; CNRS, UMR 5586 F-69622 Villeurbanne Cedex}
\date{April 30, 2008}

\begin{abstract}
We present here potential dependent mechanical properties of amorphous silicon studied through molecular dynamics (MD) at low temperature.   
On average, the localization of elementary plastic events and the co-ordination defect-sites appears to be correlated. 
For Tersoff potential and SW potential the plastic events centered on defects-sites prefer 5-fold defect sites, while for modified Stillinger-Weber potential such plastic events choose 3-fold defect sites. We also analyze the non-affine displacement field in amorphous silicon obtained for different shear regime. 
The non-affine displacement field localizes when plastic events occur and shows elementary shear band formation at higher shear strains. 
\end{abstract}

\maketitle


\section{Introduction}
\noindent Mechanical strength of crystalline and disordered solids has ever been exploited in vast research arena as well as day-to-day life. This physical property of amorphous materials like glasses, ceramics and even metallic glasses or disordered systems in general, to resist applied strain have shown immense influence in applied sciences and current technology. And the subject of improving or enhancing the mechanical strength of materials has kept its interest among researchers intact. Amorphous silicon represents an interesting choice to go beyond the study of model glasses towards the understanding of the mechanical response of real disordered systems. Amorphous silicon (from now on A-Si) is as such, distinguished from its crystalline counterpart mainly on the ground of an absence of long-range order. From the theoretical point of view A-Si is a generic example of a 4-fold covalent semi-conductor. It is also convenient to study from molecular simulations since it is a single element system for which several empirical potentials have already been developed (for e.g.,\cite{1,2}). Although a considerable amount of work has been devoted to its optical, electrical and vibrational properties over the past few years, there is still a lot to be investigated in order to understand its mechanical behaviour from the microscopic to the macroscopic scale. Previous numerical studies of the elastic and plastic response of Lennard-Jones model glasses \citep{3,4,5} have shown that at the microscopic scale the response is strongly heterogeneous and involves irreversible atomic rearrangements that appear even in the macroscopic linear regime. These systems have been studied through the analysis of the non-affine field of displacement that displays a vortex like structure. Elementary plastic events were then associated to large fluctuations of the non-affine displacement field supported by very few atoms and were observed at the boundaries between the vortices where they induce quadrupolar features in the stress field. 
         
In the present paper, we report on the elastic and plastic properties of MD generated A-Si systems at low temperatures with four classical potentials. 
Other analysis discussed later in this paper and focussed on the shear strain regime between 0 and 20\% concerns (i) the correlation between the co-ordination defects and the localization of the elementary plastic events and (ii) an analysis of the non-affine displacement field.



\section{COMPUTER SIMULATION}
Amorphous silicon structures were modelled through a molecular dynamics (MD) quench performed from a liquid phase equillibrated at 3500K. We considered here, a system of 32768 silicon atoms  with a classical potential, the Tersoff \citep{1} potential, from which model A-Si structures have already been obtained by MD quenches\citep{6}. Following the approach of \cite{6}, the Tersoff quench was done in the NVE ensemble at a density of 2.33  g$.cm^{-3}$. After reaching a temperature of about 10 K, atomic position-optimization and cell-optimization were achieved by the damped molecular dynamics algorithm in order to leave maximum total force below 10$^{-3}$ eV.\AA$^{-1}$ to release the residual stress in the system until the maximum component in the stress tensor was below  0.5 MPa.

In order to examine the dependence of the mechanical behaviour of a model A-Si on the type of classical potentials already used to describe the Si-Si interaction, we have considered apart from the Tersoff potential, three other potential forms based on the Stillinger-Weber (SW) potential: (1) the original SW potential \citep{2}, (2) a modified SW potential proposed by \cite{9} which is equivalent to the original SW potential except for the three-body term that is twice the three-body term in SW potential \citep{2}, we call it here SWM1, (3) a modified SW potential developed by \cite{10} for A-Si in which the original SW unit of energy is scaled by a factor 0.76 while the three-body term is also multiplied by a factor 1.5, we call it here SWM2.
To obtain the initial configurations of A-Si for these three SW-potentials, we started off with the Tersoff configuration generated at 10$^{11}$K.$s^{-1}$ quenching rate as mentioned above and then equillibrated it in the NVE ensemble with considered three SW potentials at 100K. Positions and cell parameters were then optimized further with the force and stress threshold similar to those used with the Tersoff potential (i.e. $f_{max}$ = 10$^{-3}$ eV.\AA$^{-1}$, $\sigma_{max}$ = 0.5 MPa).
\begin{table*}[htbp]
\begin{center}
\caption{Comparison of structural and elastic properties of A-Si system obtained with Tersoff MD quench and further relaxed with four different classical potentials, with experimental data}
\begin{tabular}{|l|c|c|c|c|c|}
\hline
{\it Unit}				&{\it Tersoff}       &{\it SW}  &{\it SWM1}&{\it SWM2}&{\it Expt.}\\ \hline
 {Density (g/cm$^3$)}	&  {2.31}            & {2.29}   & {2.20}   &  {2.24}  & {2.05\textsuperscript{a}-2.52\textsuperscript{b}}   \\ \hline
 {Average coord.}		&  {4.07}            & {4.08}   & {3.82}   & {3.88}   & {3.90\textsuperscript{c}-3.97\textsuperscript{d}}   \\ \hline
 {Average angle} 		&  {108.94}          & {108.87} & {109.22} & {109.14} & {108.6\textsuperscript{c}}  					    \\ \hline
 {Angle   dev.}  		&  {11.81}           & {11.68}  & {9.48}   & {10.40}  & {9.4-11\textsuperscript{c}}      					\\ \hline
 {B0 (GPa)}     		    &  {90.54}           & {98.55}  & {99.84}  & {76.18}  & {$\simeq$ 90\textsuperscript{e}} 					\\ \hline 
 {C44 (GPa)}     		&  {40.8}            & {34.27}  & {60.74}  & {38.61}  & {$\simeq$ 50\textsuperscript{e}} 					\\ \hline
\end{tabular}
\label{table1}
\flushleft 
\textsuperscript {a} Ref.\citep{11} $;$
\textsuperscript {b} Ref.\citep{12} $;$
\textsuperscript {c} Ref.\citep{13} $;$
\textsuperscript {d} Ref.\citep{14} $;$
\textsuperscript {e} Ref.\citep{15}\\
\end{center}
\end{table*}


\section{RESULTS AND DISCUSSION}
\subsection{Correlation between the co-ordination defects and localization of plastic events}

The structural and elastic data obtained for the calculated configurations at 0 K are compared in Table 1 with a range of available experimental data \citep{11,12,13,14,15}. The higher value of C44 that corresponds to the linear regime of shear stress vs. shear strain curves (see Figure 1) for SWM1 as compared to that for SW, is noticeable because of the factor two in three-body term that induces higher energies associated to angle distortions. For SWM2, however the similar kind of effect occurs, it is counterbalanced by the change in energy scale and results in C44 only slightly higher than the original SW model. In order to probe mechanical response, successive elementary shear strain of step-size $\delta$$\varepsilon$ = 10$^{-3}$ was homogeneously applied in direction of step-wise incremental strain on the obtained four A-Si model configurations by small variation in the lattice parameters. After each deformation, the structure was relaxed until the maximum total force in the system was brought below 10$^{-3}$ eV.\AA$^{-1}$. The stress-strain curves obtained with the four empirical potentials in this study are shown in Figure 1. In the low strain regime the stress curves for the four potentials closely follow the linear behaviour. The highest slope is found for the SWM1 potential and the lowest for the SW potential which corresponds to their elastic constants (see Table 1). This initial linear regime observed in the amorphous samples is not necessarily associated to an elastic behaviour as the atomic structure can experience some local irreversible rearrangements associated with plastic events. The break-down point, often called as a yield point, refers to the situation when system no longer remains stable under constant shear stress and undergoes a macroscopic or a large-scale deformation to achieve the stable equillibrium state. 
A comparison of the behaviour of the A-Si configurations studied with the SW models in the break-down region shows a strain variation of 15\% on the shear strain and 145\% of variation on the shear stress at which the break-down occurs i.e. to say that yield stress by a large amount is more sensitive to these potentials than yield strain. Unlike the SW models, the Tersoff model does not see a large break-down in the shear stress in the applied shear strain region.         

\begin{figure}[htbp]
\centering
\includegraphics[width=2.5in,height=1.6in]{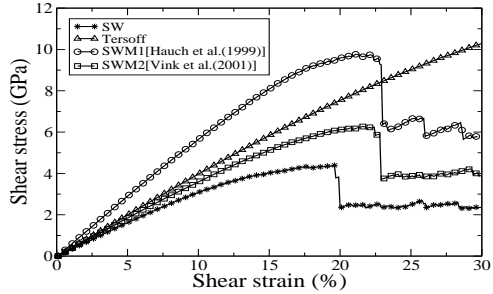}
\caption{Shear stress vs. shear strain for four A-Si configurations}
\label{figure1}
\end{figure}

Quasi-static shear strain applied successively to A-Si models obtained with four potentials causes the re-arrangement in their structures and hence, the change appears in the number of nearest neighbors or co-ordination defects. The evolution of four fold co-ordination (\%) with applied shear strain (\%) is plotted in Figure 2. For all four potentials the number of 4-fold co-ordination decreases with applied shear strain and defects created are very few for the Tersoff potential even for shear strain as large as 10-20\%, while it is large for Stillinger-Weber (SW) potential. A sharp increase in the number of defects is seen around 19.9\% shear strain for SW potential (by the amount 4-fold co-ordinated atoms decrease). The difference between the SW, SWM1 and SWM2 potentials can be rationalized by higher values of 3-body term in modified SW potentials that tends to hamper formation of 5-fold atoms and favours lowering the energy for 3-fold defects instead. During deformation,predominantly 5-fold defects are created for Tersoff and SW potentials, while for SWM1 and SWM2 potentials mostly 3-fold defects are created.    

\begin{figure}[htbp]
\centering
\includegraphics[width=2.5in,height=1.6in]{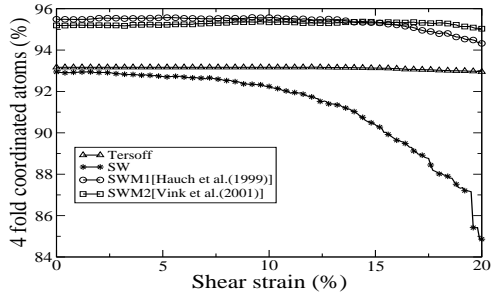}
\caption{Four-fold co-ordination (\%) vs. shear strain (\%)}
\label{figure2}
\end{figure} 

In order to evaluate the correlation between the localization of the plastic events and the co-ordination defects, it is necessary first to identify plastic events in the modelled A-Si systems. As a first step, the coarse-graining of the non-affine displacement field was implemented at each successive shear strain step. 
For this we used a procedure detailed in \cite{5}, where the non-affine displacement field on the particle is written as : 
$ U^{NA}_i  = || {\bf r^{n+1}_i} - {\bf r^{n}_i} - {\bf \delta_{aff} r^{n}_i} || $ , where ${\bf r^{n+1}_i}$ represents positions of particle $i$ in  ${(n+1)}^{th}$ strain-configuration (for e.g, configuration for shear strain $\varepsilon = (n+1)\cdot 10^{-3}$ ),  ${\bf r^{n}_i}$ represents positions of particle $i$ in the ${n}^{th}$ strain-configuration (for e.g., $\varepsilon =  n \cdot 10^{-3}$ ) and ${\bf \delta_{aff} r^{n}_i}$ represents the affine displacements of particle $i$ from configuration $n$ to $(n+1)$ calculated as $ {\bf \delta_{aff} r^{n}_i} = {\delta \bar {\bar \varepsilon}} \cdot {\bf r^{n}_i}$. Here,${\delta \bar {\bar \varepsilon}}$ is the homogeneous strain tensor associated with elementary shear strain step ${\delta \varepsilon}$.         

The amplitude of non-affine coarse-grained (CG) displacement field was then written on a regular grid in order to apply an attractor analysis that is closely related to the Bader charge analysis \citep{16,17} used in Quantum Chemistry. This CG amplitude of displacement field on a regular grid is calculated using the following expression : 
$f({\bf r_g}) =  {\large \sum}_i  (U^{NA}_i)^2
 ({ 1\over{ \sqrt{\pi} \omega}  })^{3} e^{-  {||{\bf r_g} - {\bf r^n_i}||^2 \over \omega^2}}$, where  ${\bf r_g}$ are positions of the grid points and $\omega$ is the characteristic gaussian width (2.55 \AA) which approximately corresponds to the Si-Si first neighbour distance.   

Following the definitions of the Bader charge \citep{17}, we then associated attractors to the maxima of $f({\bf r_g})$ and determined their basins by finding the volumes defined by  $\vec {\nabla} f \cdot \vec{n_S} = 0 $ where $\vec{n_S}$ is the normal to the surface of the basin.   
Practically the calculation of the attractors and basins are done on a discrete grid (100$\times$ 100 $\times$ 100) by using a simple scheme based on finite differences which has already been used in  context of atomic charge calculations \citep{18}. In that scheme, points on a grid belong to one or more attractors with some weighting factors. 
In the most simple approximation that we used here, the weighting factors are simply $ 1 \over n_{att} $, where  $n_{att}$ is the number of attractors the grid point belongs to.         
For each attractor we calculated the associated basin as well as the sum of $f({\bf r_g})$ over this volume, which we will refer from now on, as the integrated intensity.   

Typical distributions of the integrated intensities for two different strain values are shown in Figure 3 for the Tersoff potential case. 
The first curve (Figure 3(a)) shows a continuous distribution while a large gradient is observed around the most important attractor on Figure 3(b). 
In this last case we associated a plastic event to this large attractor which also bears the highest local value of the non-affine field and the larger basin. 
Following the same procedure for all the strain-configurations, we used such large variations (more than $1.5 \times 10^{-4} \AA^2$ ) in the distribution of integrated intensities to define local plastic events. 
Along with this criterion, we also define local plastic events when atoms closer than 3.4 \AA\ from an attractor display a non-affine displacement larger than 0.05 \AA.

\begin{figure}[htbp]
\centering
\includegraphics[width=2.5in,height=1.5in]{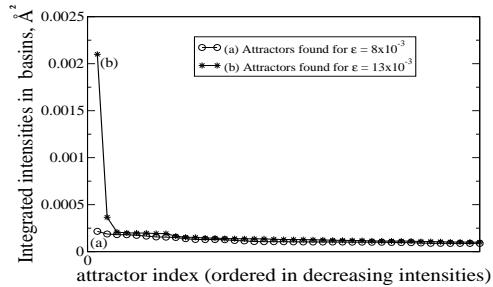}
\caption{Distribution of the integrated intensities (see text) for A-Si obtained with the Tersoff Potential}
\label{figure3}
\end{figure}  
 
Now, to check the quality of the plastic event definition we applied a quasi-static shear-strain with $\delta \varepsilon = -10^{-3}$ step-size to all the $(n+1)$ strain-configurations. We then obtained the residual displacement by calculating the difference in positions with the $n$ strain-configurations. For a perfectly elastic system this residual displacement is exactly zero, while a non-zero value is expected if plastic events have occured.  

In Figure 4, we compare the square of maximum residual displacements with the integrated intensities summed over detected plastic events. Whenever a large maximum residual displacement is found, it systematically associates with one or more detected attractors, which demonstrates an efficiency of this technique to identify the local irreversible (plastic) events. Only those maximum displacements lower than 0.007 \AA\ are not detected as plastic events. Although the magnitude of the maximum displacement and the basins intensities are not perfectly correlated, a good qualitative agreement is evident. This suggests that the intensities of the detected attractors may be used to obtain an evaluation of the plastic activity in the system.

\begin{figure}[htbp]
\centering
\includegraphics[width=2.5in,height=1.6in]{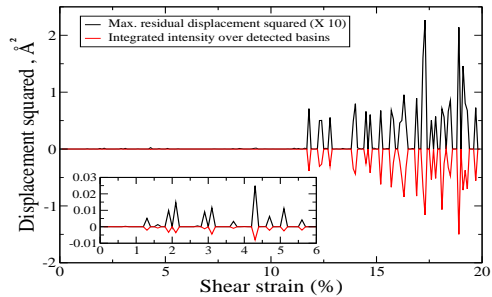}
\caption{Comparison of the square of maximum residual displacements with the integrated intensities summed over detected plastic events}
\label{figure4}
\end{figure}


Figure 5 shows the amount of detected plastic events that localizes on co-ordination defects or on sites which are first neighbours of co-ordination defects. Practically, we sought out the co-ordination number of atom in close proximity to each localized attractor on a regular grid.
The co-ordination number is calculated by summing all the neighbours of this atom in a shell of 2.85 \AA,  which is approximately ($\pm$ 0.05\AA) the position of second minima in the Si-Si radial distribution function. 
Figure 5(c) represents expected proportion of plastic events that would localize on defect sites or on their neighbours if the plastic events were distributed with a uniform probability over all the possible sites. 
The lower and upper diamond-dashed curves (c) in Figure 5 respectively correspond to the zero strain and 20\% strain configurations which is intuitive from the fact that the number of defects increases with the shear-strain deformation.
The calculated distributions (Figure 5(a),(b)) always correspond to a defect localization ratio that is higher than for the uniform probability distribution (Figure 5(c)). We can, therefore conclude that there exists a positive correlation between the localization of the plastic events and the defect sites. Depending on the potential this correlation can be more or less marked. In case of the Tersoff potential almost all the plastic events  localize on the defect sites or on their neighbours (Figure 5(a)), while for the SW potential it is true for 63\% of the plastic events. These findings are in good agreement with the analysis of \cite{7}, where it is stated that defect sites in a model Stillinger-Weber A-Si could play a role of plasticity carriers.

A slightly more detailed analysis concerns the nature of the defect sites on which the plastic events localize. From Table 2, it can be noticed that the type of defect on which the plastic events localize is largely potential dependent. The localization mainly occur on 5-fold co-ordinated defects for SW and Tersoff potentials while it is found mainly on 3-fold defects for SWM1 and SWM2. In their analysis, \cite{7} assign the plastic activity to the presence of liquid-like silicon atoms that in context of the present study is to be related with the 5-fold co-ordination defects. Our results show that the specific role of these liquid-like silicon atoms is mainly due to the details of the Stillinger-Weber potential rather than to a property of the amorphous silicon systems. However,our results on different types of potentials indicate that the type of defects on which the plastic events tend to localize also correspond to the type of defects that are preferentially created during the application of the shear strain deformation.

\begin{table*}[htbp]
\begin{center}
\caption{Plastic events localization on type of defect sites for considered classical potentials}
\begin{tabular}{|l|c|c|c|c|}\hline
\it{Potentials} & \multicolumn{4}{c|}{\it{Type of defects sites on which plastic events localize}}\\
\cline{2-5}
               & {\it{2-fold (\%)}}  & {\it{3-fold (\%)}}  & {\it{5-fold (\%)}} &{\it{6-fold (\%)}} \\ \hline 
      Tersoff  & {0}    		        & {0}   		      & {97.8}	            &{2.2}	    \\ \hline 
      SW       & {0}    	                & {6.8}    		      & {90.6}	            &{2.6}		    \\ \hline  
      SWM1     &	 {5.3}     	        & {83.4}    		      & {11.1}	            &{0}	    \\ \hline 
      SWM2     &	 {2.5}    	        & {70.0}    		      & {27.5}		        &{0}	    \\ \hline  
\end{tabular}
\label{table3}
\end{center}
\end{table*}


\begin{figure*}
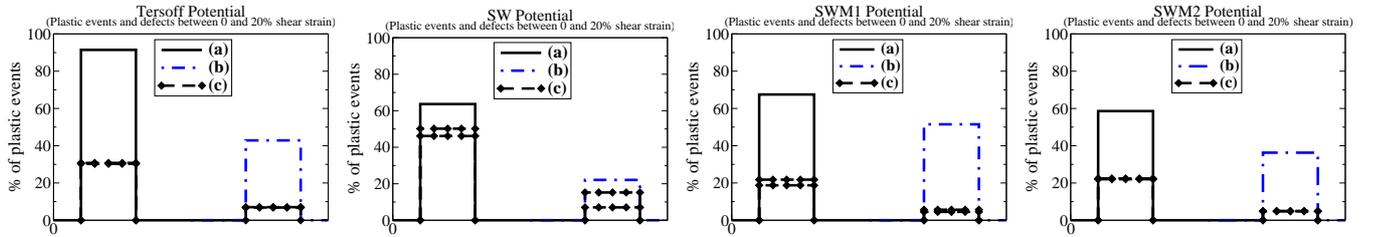

  \begin{center}
    \begin{tabular}{cccc}
      \resizebox{1.7in}{!}{\includegraphics{graph1cTER_1.eps}} &
      \resizebox{1.7in}{!}{\includegraphics{graph1cSW_1.eps}} &
      \resizebox{1.7in}{!}{\includegraphics{graph1cSWM1_1.eps}} &
      \resizebox{1.7in}{!}{\includegraphics{graph1cSWM2_1.eps}} \\
    \end{tabular}
    \caption{Plastic events localization: (a) on defect sites or on their first neighbours (b) on the defect sites only (c) ratio of plastic events localized on respective defect sites (a) or (b) that would be obtained from a uniform probability distribution}    
    \label{figure5}
  \end{center}
\end{figure*}


\subsection{An analysis of the non-affine displacement field} 

The stress follows shear strain linearly upto the regime where non-affine displacement is irreversible. Then it departs slightly from linearity entering the zone of irreversibility and localization of non-affine displacements. Beyond the breakdown point strongly localized irreversible displacements induce the formation of elementary shear bands (see Figure 6). Figure 6 shows for A-Si modelled with SW potential, the pattern of the coarse-grained non-affine displacements in distinct regimes of applied shear strains range, namely the region corresponding to reversible displacements, the region where non-affine displacements localize and the region corresponding to elementary shear band formation. The maximum non-affine displacements at the centre are around 1000 times smaller than the total displacements at boundaries of the A-Si cell when there are no plastic events. For Figure 6(a), \(\delta U_{NA} \simeq 1\times 10^{-4} \cdot \delta U_{ybound}\), where \(\delta U_{ybound}=2\varepsilon_{xy} \cdot L_{x}\), \(\varepsilon_{xy}\) being the shear strain applied along Y-axis; $L_{x}$, the length of cell of A-si along X-axis; \(\delta U_{NA}\), the non-affine displacement field and \(\delta U_{ybound}\), the total displacement field at the boundaries of the A-Si cell along Y-axis. 
Note that the non-affine displacement field was \(\sim 0.01\times \delta U_{ywall}\) for Lennard-Jones (LJ) glass \citep{3}. It means that here the non-affine displacements are 100 times smaller than in LJ systems. Quadrupolar like structure in non-affine displacement field associated with the elementary plastic events in the system as shown in Figure 6(b) corresponds to the case when maximum non-affine displacements at the centre are approximately few tens of inter-atomic distance, $d_{ia}$ i.e. \(\delta U_{NA} \simeq 0.1\times d_{ia}\) and Figure 6(c) describes the case of elementary shear band formation, where \(\delta U_{NA} \simeq \ d_{ia}\).  
In general, for all potentials, non-affine displacement field is inhomogeneous in reversible region but spans the entire system size and they are localized in case of occurrence of plastic events.     


\begin{figure*}
  \begin{center}
   \begin{tabular}{ccc}
      \subfigure[a]{
      \label{fig:subfig:a}
      \resizebox{1.7in}{!}{\includegraphics{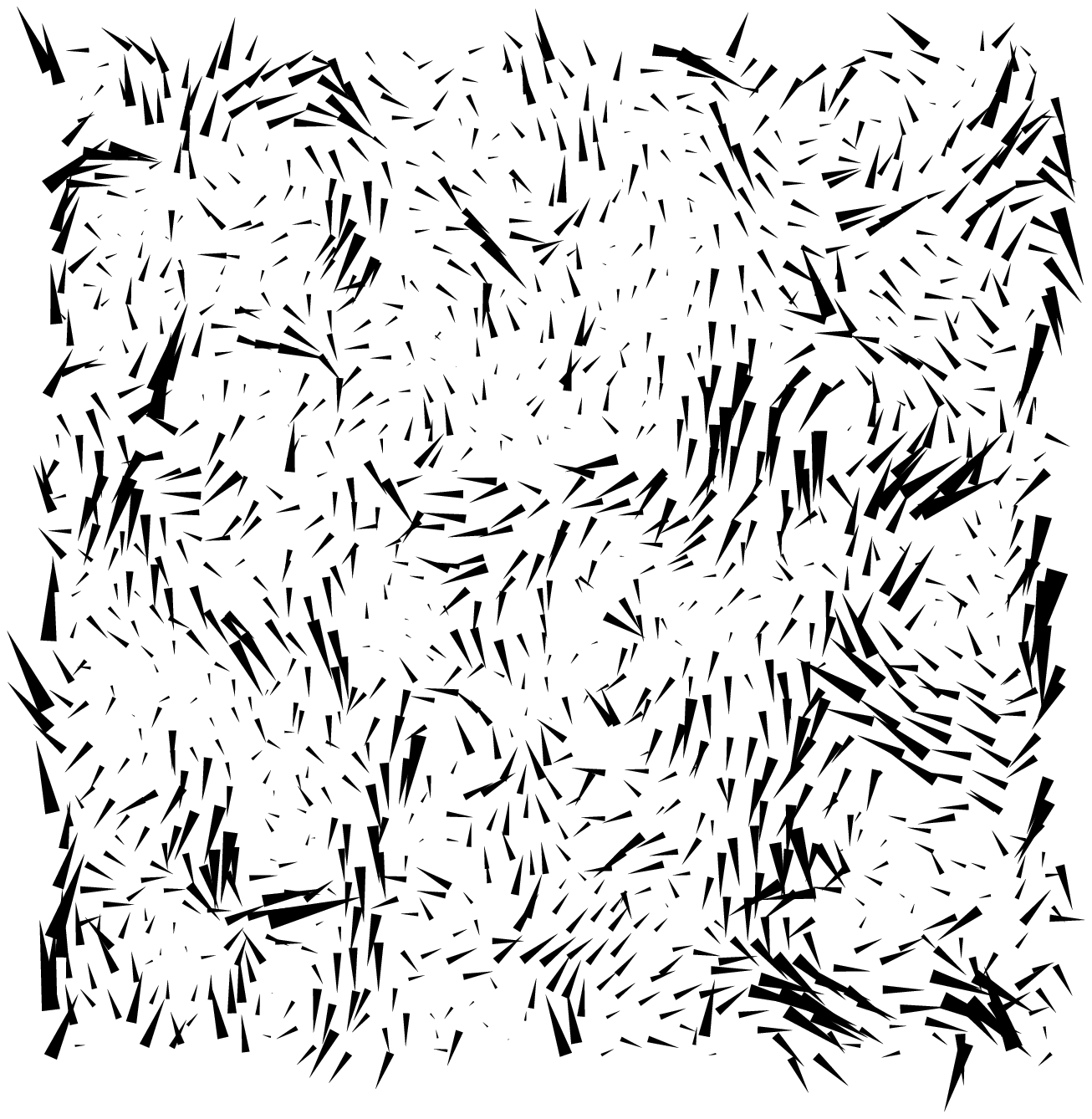}}}  &
          \subfigure[b]{
          \label{fig:subfig:b}
          \resizebox{1.7in}{!}{\includegraphics{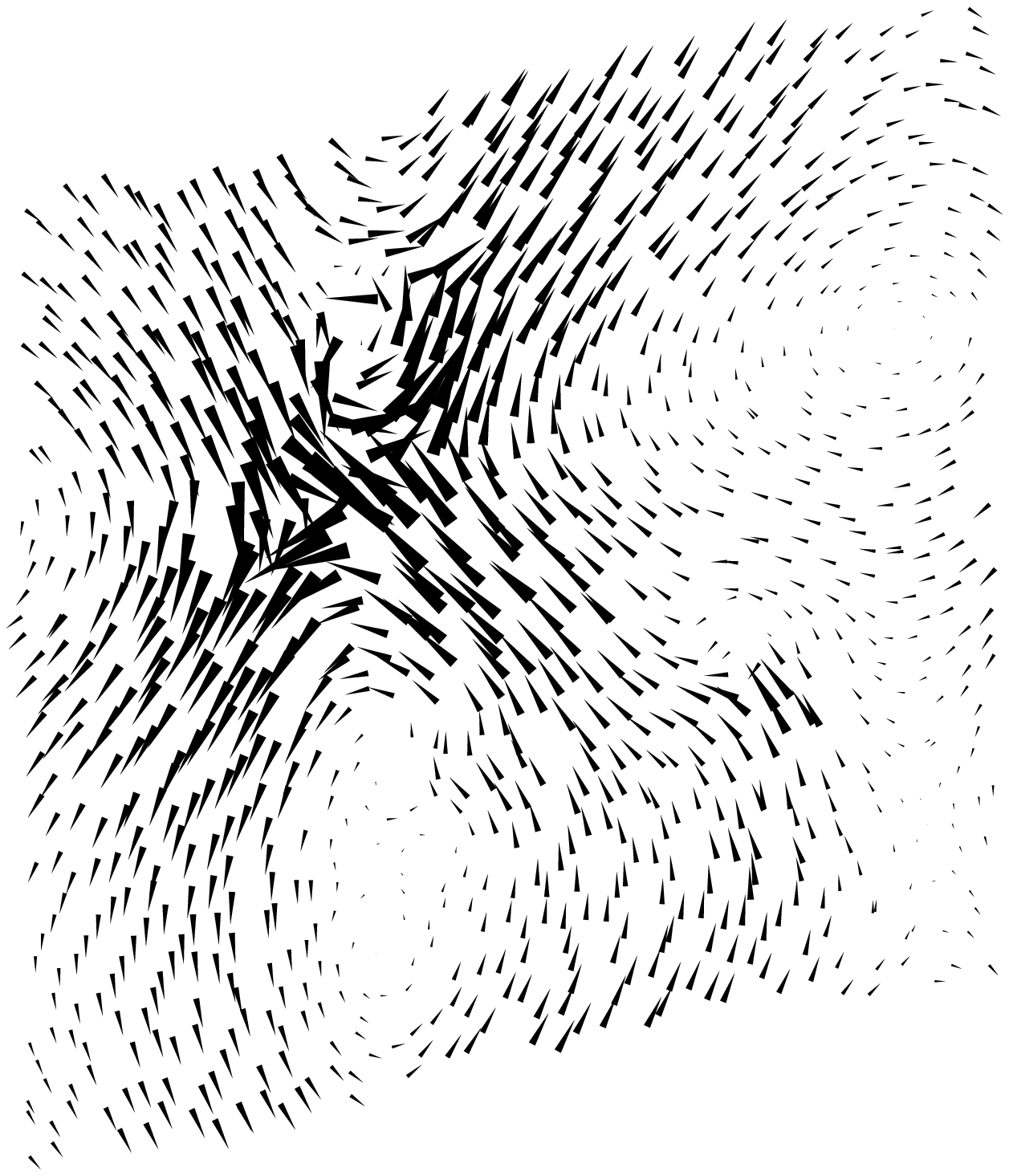}}} &
      \subfigure[c]{
      \label{fig:subfig:c}
      \resizebox{1.7in}{!}{\includegraphics{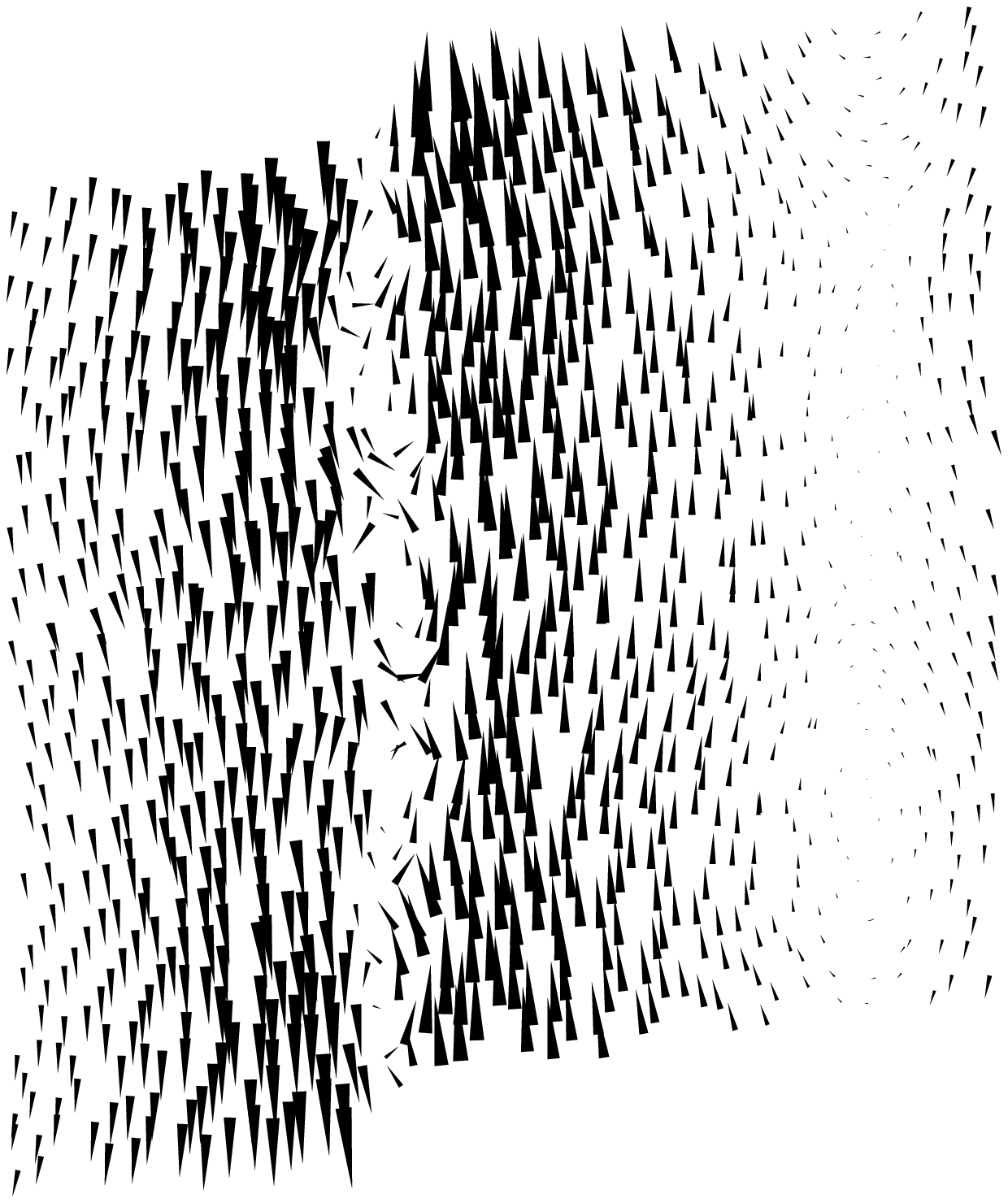}}}
    	 \end{tabular}
    \caption{Coarse-grained non-affine displacement fields for A-Si modelled with SW potential in the shear strain range between 0 and 20\%}
   \label{figure6}
  \end{center}
\end{figure*}


\section{CONCLUSION} 

The stress-strain relation, however shows in general similar behaviour, is potential dependent. 
For the Tersoff and the SW models, the plastic events localize mainly on 5-fold co-ordinated defects, while 3-fold co-ordinated defects are mainly preferred for SWM1 and SWM2. The non-affine displacement field shows quadrupolar like structure when there are plastic events and forms elementary shear bands for higher strain values for SW potential models which exhibit a large stress drop around yield strain,as in other model amorphous solids without local tetragonal order. Only the amplitude of the non-affine contribution to the displacement field vary which is much lower for silicon than for other systems.  



\begin{acknowledgments}
The authors would like to acknowledge the financial support by ANR-PlastiGlass, France and computational support by IDRIS/France, CINES/France and CEA/France. We also thank Prof.~J-L.~Barrat, Prof.~B.~Champagnon and Prof.~D.~Vandembroucq for stimulating discussions.
\end{acknowledgments}


\end{document}